\documentclass[12pt]{spie} 
\usepackage{graphicx}
\usepackage{amssymb} 
\usepackage{amsmath}    
\usepackage{latexsym}   
\usepackage{bm}
\usepackage{times}

\def\half{\textstyle{\frac{1}{2}}}

\title{Clocking in the face of unpredictability \\beyond quantum
  uncertainty}

\author{F. Hadi Madjid\supit{a} and John M. Myers\supit{b}
\skiplinehalf
\supit{a}82 Powers Road, Concord, MA 01742, USA;\\
\supit{b}Harvard School of Engineering and Applied 
Sciences, Cambridge, MA 02138, USA}

\authorinfo{Further author information: (Send correspondence to J.M.M.)\\J.M.M.: 
E-mail: myers@seas.harvard.edu, Telephone: 1 617 495 5263\\  
F.H.M.: E-mail: gmadjid@aol.com, Telephone: 1 978 369 1808}

\begin{document}
\maketitle


\begin{abstract}
In earlier papers we showed unpredictability beyond quantum uncertainty in
atomic clocks, ensuing from a proven gap between given evidence and
explanations of that evidence.  Here we reconceive a clock, not as an isolated
entity, but as enmeshed in a self-adjusting communications network adapted to
one or another particular investigation, in contact with an unpredictable
environment.  From the practical uses of clocks, we abstract a
clock enlivened with the computational capacity of a Turing machine, modified
to transmit and to receive numerical communications.  Such ``live clocks'' phase the
steps of their computations to mesh with the arrival of transmitted numbers.
We lift this phasing, known in digital communications, to a principle of
\emph{logical synchronization}, distinct from the synchronization defined by
Einstein in special relativity.  

Logical synchronization elevates digital communication to a topic in physics,
including applications to biology.  One explores how feedback loops in clocking
affect numerical signaling among entities functioning in the face of
unpredictable influences, making the influences themselves into subjects of
investigation.  The formulation of communications networks in terms of live
clocks extends information theory by expressing the need to actively maintain
communications channels, and potentially, to create or drop them.

We show how networks of live clocks are presupposed by the concept of
coordinates in a spacetime. A network serves as an organizing principle,
even when the concept of the rigid body that anchors a special-relativistic
coordinate system is inapplicable, as is the case, for example, in a generic
curved spacetime. 
\end{abstract} 

\section{Introduction}
Discoveries in physics involve unpredicted events, whether of the imagination of
theorists or the surprises of unpredicted experimental evidence.  Still, one
can wonder: if scientists were smarter, could they have predicted the
events that have come as surprises?  Within the framework of quantum
theory, the answer is ``not on the basis of experimental evidence.''  As is
explained in Sec.\ \ref{sec:2}, quantum explanations depend on guesses that are
subject to revision in the face of future evidence, not only as a practical
matter but as a matter of principle, demonstrating an inescapable
unpredictability beyond the probabilistic uncertainty of quantum mechanics.

The proven dependence of explanations on guesswork promotes the aspiration,
never quite attainable, to separate explanations from evidence, and to attend
to the sources of both.  I find a path to discovery when I recognize that my
explanation cannot be the only one possible.  This recognition of a certain
freedom in choosing explanations may be important to philosophy, but does it
matter to physics?  To begin with, because investigations draw on surprises,
both from external outcomes and from acts of our own guessing and making of
assumptions, we who investigate are not interchangeable: You experience what I
do not, and I can know what you find out only to the extent of our
communication.  Today investigators are often linked to sensors, to actuators,
and to each other by a computer-mediated communications network.  Experimental
activity, including encounters with unpredictability, is reflected in such a
network, both in the records held in the computers of the network and in the
timing of communications within the network.  Networks involve
feedback. Indeed, lots of biological and engineered mechanisms involve networks
that respond to unpredicted deviations from some aiming point.  We want to lift
the feedback that takes place in a computer-mediated system from the province
of engineering to a form suitable for application in theoretical physics.

As an application of feedback in computer-mediated networks, we consider
investigations of motion.  The motion of an object can be thought of or
measured only in relation to organized motions of other objects that serve as a
reference background against which to see the motion of the object under
investigation.  Here we argue that recognizing the gap between evidence and its
explanations forces a reconception of reference motions, away from the a rigid
system of equations and measurement units, conceptually tied to a single
investigator, to a pattern of motion maintained by investigators and their
devices, linked in some chosen communications network that the investigators
create and maintain, operating in an environment of surprise and guesswork.

To develop the concept of a networked reference for motion we start by
retracing the topic, opened up a century ago by Einstein, of spatially and
temporally locating events of interest.  At root, this is a problem for every
person from infancy on.  Right now I feel my fingers on the keyboard, the reach
of my arm for the coffee cup, the chair I sit in; I see the walls of the office
and, through the window, the bare trees and the snow lingering in this
late-coming season of spring.  Or think of Galileo comparing the motion of a
rolling ball against himself, e.g. his own heart beat; then as science
progresses he (and we who follow after him) delegate the beats of our reference
to a more stable external clock.  A clock and a flat floor make a background
for motion; however, reading a distant clock is an issue.

The prevalent way to think of a background against which to define motion is a
coordinate system, but the abstraction that generates a coordinate system
obscures the reference motions that necessarily compose such a background.  To
recover the reference motions intrinsic to a coordinate system, we review a
little history.  Today's notion of a coordinate system stems from the theories
of special and general relativity.  Commenting on how he developed those
theories, Einstein spoke of the need to ``drag down from the Olympian fields of
Plato the fundamental ideas of thought in natural science, and to attempt to
reveal their earthly lineage \ldots'' \cite{einstein61}.  The earthly lineage
leading to the theory of special relativity lies in the railroad age of the
1800's, in which the development of telegraphy provided the means to
synchronize clocks to tell ``time'' spread across the span of telegraphy.
Einstein abstracted the practice of 
synchronizing clocks by the exchange of telegraph signals into the theory of
special relativity.  Building on his bold hypothesis about light speed, he
defined the synchronization of imagined proper clocks in terms of clock
readings at the transmission and reception of idealized light signals.  He defined
a coordinate frame relative to an imagined rigid body in terms of signals
exchanged among synchronized proper clocks fixed to the body.  This definition
tied the concept of a rigid body to a pattern of signal exchange among proper
clocks, thereby relativizing the notion of \emph{rigid}.  Without clocks and
the motion of the signals by which their synchronization is defined, there can
be no concept of a Lorentz coordinate system.

In general relativity, no extended body can be rigid, but the equations that
define a metric tensor field are still imagined as fixed.  For astronomy and
geodesy, the International Astronomical Union (IAU) employs a notion of a
`reference system' incorporating fixed equations defining a metric tensor field
\cite{soffel03}.  Although obtained by dragging down ideas ``from the Olympian
fields of Plato,'' Einstein's synchronization criterion for special relativity,
as well as the general-relativistic reference systems, are elevated to a
blackboard of Plato, suitable for armchair contemplation, but remote from any
active experimental setting.  From the perspective of a physics that faces the
unpredictable, the blackboard on which equations are written needs to be split
up and made dynamic, with the equations of a reference system copied into the
separate memories of the computers of the network that links the investigators
and their automated devices.  Copied into these memories, the equations become
available to distributed real-time computations essential to the feedback that
maintains the network; furthermore, on occasion, the equations written in one
or another computer memory can be modified in response to unpredictable
happenings.

What can be investigated depends in part on how the computers of a network can
communicate.  In Sec.\ \ref{sec:3} we represent a real-time process-control
computer that takes part in a network as a modified Turing machine stepped by a
clock, and we explore the form of synchronization, distinct from Einstein's,
necessary to the transmission of numbers from one Turing machine to another.
To deal with communications between Turing machines, it will be necessary for
the clock that steps a Turing machine to tick at a rate that can be adjusted by
commands issuing from that machine.  We call such an adjustable clock in
combination with the Turing machine that regulates its rate a \emph{live
  clock}.  (Such rate adjustment is found in the cesium clocks that generate
International Atomic Time (TAI)\cite{aop14}.)  Because the evidence recorded by
one live clock differs unpredictably from that recorded by another, unlike
proper clocks postulated in relativity theory, each live clock is unique.

A live clock can receive a number from a detector, such as a photodetector,
that acts from a computational point of view as an oracle, the workings of
which are unknowable\cite{turingTh}.  Other oracle numbers show up as clock
readings made by one live clock and transmitted to another live clock.  Both
human and automated communication of numbers involves cycling through phases.
Think of the cycle of a bucket brigade, in which a person turns to receive a
bucket just as a another person offers the bucket.  This need for cycling
through phases structures the communications in which live clocks participate,
leading to a definition of the \emph{logical synchronization} of a channel from
one live clock to another, distinct from the synchronization defined by
Einstein.  The need for logical synchronization of its channels constrains the
design of a network of live clocks.

To maintain logical synchronization of the channels of a network of live clocks
requires feedback because, among other reasons, at the level of precision with
which atomic clocks can be measured, ``no two clocks tick alike.''\cite{aop14}
In this feedback, the live clocks of a network steer their tick rates and
accelerations in response to unpredictable measured deviations from an aiming
point.  The aiming point is chosen to express desired communications channels,
and steering toward that aiming point is determined by control equations
programmed into the live clocks.

Designing and operating the channels of a network and the corresponding control
equations requires a hypothesis concerning how signals propagate from one live
clock to another.  It is a great convenience in making such a hypothesis to
draw on the concept of a spacetime manifold; however, evidence in the form of
clock readings is independent of any hypothesis about signal propagation, and
therefore can be used to test a hypothesis, regardless of whether the
hypothesis invokes the concept of a spacetime manifold.  Thus there is great
freedom in the choosing of hypotheses about signal propagation.  In addition to
serving as a background against which to think about and measure motion, a
network of live clocks serves as a tool to investigate the propagation of the
signals by which the live clocks of the network communicate.

Among candidate hypotheses about propagation, those based on a
general-relativistic spacetime manifold commend themselves as an appropriate
first case for theoretical purposes.  It is especially interesting that a
generic curved spacetime manifold rules out a fixed general-purpose pattern of
signal exchange, and so calls for tailoring a network of live clocks to
whatever situation is under investigation.  In Sec.\ \ref{sec:4} we discuss a
network of live clocks, designed assuming a spacetime manifold and intended to
serve as an adaptive background for motion and as a tool to investigate
gravitation.  The operation of the network involves a cycle of provisionally
assuming a metric tensor field, using it to construct an aiming point in terms
of relations among clock readings toward which to steer the live clocks,
implementing the corresponding control equations, and evaluating the resulting
deviations from the aiming point. If the deviations cannot be held within
tolerable bounds, the investigators, drawing on the measured clock readings,
guess a different metric tensor field, and the cycle continues.

Sec.\ \ref{sec:5} offers a brief perspective on how recognizing essential
unpredictability impacts the application of physical concepts, based on the
dependence of both explanations and experiments on a dynamically evolving
``tree of assumptions.''

\section{Unpredictability implied by quantum theory}\label{sec:2}
Given a quantum state and a measurement operator, quantum theory sets up
predictions of a probability measure over a set of possible outcomes, and
quantum uncertainty denotes a spread in the probability measure.  Quantum
theory, however, implies something beyond uncertainty, namely a kind of
unpredictability, in some cases involving the introduction of possibilities
previously unforeseen.  In earlier work we proved that:
\begin{enumerate}
\item Infinitely many explanations are consistent with any given evidence.
  While each explanation fits the given evidence, the explanations have
  conflicting implications for evidence obtainable from experiments not yet
  performed. Once candidate explanations are proposed, one can bet on them---in
  effect, assigning rough probabilities to them.  But in the face of an
  experimental surprise, before any explanation is proposed, no evidence can
  determine the probability to assign to any of the numberless not-yet-known
  explanations that fit given evidence.  For this reason, explanations
  as hypotheses chosen for testing come as surprises, unpredictable even to
  the person who chooses them.  \cite{ams02,aop05,CUP}.
\item
To choose one or a few explanations requires reaching beyond any predictive
logic and beyond the guidance of probabilities to make a guess \cite{aop05}.
\item Because it excludes conflicting explanations that fit the given evidence
  equally well, any guessed explanation is susceptible to refutation by
  surprising results coming from future experiments, prompting its replacement
  by some other explanation \cite{tyler07}.
\item Choosing explanations as hypotheses to be tested by experiments takes
  place in an open cycle, alternating between subjective surprises of guessing
  hypotheses and objective surprises of unexpected experimental results
  \cite{tyler07}.  Thus not only do explanations depend on assumptions but so do
  experiments: their design and the interpretation of their results depend on
  explanations that rest on layers of guesswork.
\item In addition to the unpredictability of guesses that choose explanations,
  quantum uncertainty makes individual occurrences of outcomes of generic
  measurements unpredictable.  Two situations are to be distinguished:
  \begin{enumerate}
  \item An experimenter tests a predicted probability measure by repeating a
    measurement many times and comparing measured relative frequencies against
    predicted probabilities.  In this situation individual occurrences of
    outcomes matter only as they enter tallies, ratios of which become relative
    frequencies.
\item As in applications of quantum decision theory \cite{helstrom,holevo}, an
  experimenter has a prior probability measure, sometimes only a guess, for the
  state being measured and uses a single unpredictable result or a short run of
  results to together with Bayes rule to decide on a parameter value that
  characterizes the state, acting promptly on that decision.  (Such prompt
  action takes place in the feedback loops by which atomic clocks must be
  steered in frequency \cite{aop14}.)
  \end{enumerate}

\end{enumerate}
Although surprises in physics are familiar as a practical matter, before the
proof one might try to relegate them to the status of exceptions to the
progress of science.  The proof gives theoretical force to the unpredictability
of surprises and guesses, so that the implications of this unpredictability can
be explored.  If a guess is necessary for an explanation, what does that mean
for discovery?  Who or what makes a guess?  In particular, the dependence of an
explanation of given evidence on guesswork motivates more attention to two
things: (a) the distinction between explanations and the experiences that one
explains, and (b) the entity, long neglected in physics, in which evidence,
calculation, and guesswork meet.

\section{Clocks that compute and communicate}\label{sec:3}
When investigators and their devices are linked to one another by a
computer-mediated network, the network serves as a tool for the investigation;
it also houses the results of the investigation in records distributed in the
memories of its computers.  We model a computer in a network by a clock joined
to a Turing machine that is modified to permit communication with other such
machines and with an unpredictable environment containing sensors such a
photodetectors.  The clock has provision for its rate of ticking to be adjusted
on the fly by commands from the Turing machine.  We call such a clock together
with the Turing machine a \emph{live clock}.  The live clock models an entity
in which evidence, guesswork, and calculation meet.  It is the key component in
any of various communications networks against which to reference the motion of
objects under investigation.  Live clocks enable a networked reference for
motion to adapt to the needs of particular investigations.

Turing represented mathematically the activity of calculating by what is now
called a Turing machine, imagined as operating in a sequence of `moments'
interspersed by `moves'.  The Turing machine has a working memory in the form
of a tape.  At any moment the machine scans one square of the tape, on which it
can read or write a single character of some alphabet that need not be more
than the binary set $\{0,1\}$ \cite{turing}.  A move as defined in the
mathematics of Turing machines consists only of the logical relation between
the machine at one moment and the machine at the next moment.  Thus the Turing
machine models the logic of computation by relating a program and a record in
memory to an output in memory, without regard to timing. Two computations
executing at different speeds can be represented in their logic by the same
sequence of moments and moves.  Because the Turing machine is indifferent to
timing, it cannot calculate the tick that steps its moves, and so cannot by
itself express the physical motion of computation in the way that the live
clock does.

People or machines as calculators do lots of things, such as taking a break or
fetching new supplies, that Turing abstracted out of sight.  In so doing he
sealed the Turing machine off from receiving or transmitting communications to
anything outside its computational activity.  He did, however, open the door to
communication a crack by briefly discussing a variant that he called a
\emph{choice machine} which, on occasion, could wait for an external input
\cite{turing}.  In adapting it as a component of a live clock, we modify the
Turing machine further so that the live clock can not only receive characters
of its alphabet from an external environment, but can also transmit characters
to that environment.  The environment can include other live clocks.

\subsection{Cycle of moves and moments of computation}
A live clock operates in a cycle of receiving unpredictable information from an
environment, storing that information in memory, computing a response, and
issuing that response to the environment.  The cycle has subcycles, and at the
finest level is composed of moments and moves of the clock-driven Turing
machine that makes up the live clock.  For a live clock to take part in
communication, its moments and moves have to be regulated to avoid the logical
conflict of a collision between writing into memory and reading from
memory. (In human terms this is the collision between trying to speak and
listen at the same time.)  To avoid this conflict,  the modified Turing machine is
driven by the adjustable clock through a cycle with two phases of moves and two
phases of moments, with reading from memory taking place in a phase separated
from a phase of writing into memory.

A cycle of the live clock corresponds to a unit interval of the readings of its
adjustable clock.  A reading $t$ of a live clock can be expressed in the
form $m.\phi_m$ where an integer $m$ indicates the count of cycles and $\phi_m$
is the phase within the cycle.  We choose the convention that $-1/2<\phi_m\le
1/2$. (It is not necessary to think of the signals as points in time; it
suffices to think of a point reference within the signal.)

Guesses, such as guessed explanations, enter the memory of a live clock as
inputs from outside of the live clock, and are written into its memory during
phases of writing.  We give no further explanation of this ``outside'' from
which guesses are assumed to come.  Because live clocks have memories, they can
record and make use of readings of other live clocks and indeed their own
readings---a mechanical analog of ``self-awareness.''

  \subsection{Logically synchronizing the communications between live clocks}
To express the clocking of actual or contemplated communications between one
live clock and another, we follow Shannon in speaking of a communications
channel; however we augment his information-theoretic concept of a channel
\cite{shannon48} with the live-clock readings at the transmission and reception
of character-bearing signals \cite{aop1 4}.  Each character transmitted
from a live clock $A$ to a live clock $B$ is associated with a reading of live
clock $A$ of the form $m.\phi_m$ at the transmission and with a reading of live
clock $B$ of the form $n.\phi_n$ at the reception.  A channel from $A$ to $B$
includes a set of such pairs of readings of the transmitting and the receiving live
clocks.  The necessity of avoiding a conflict between reading and writing
imposes a constraint on the phases of reception.
\begin{quote}
\textbf{Proposition:} A character can propagate from one live clock to another
only if the character arrives within the writing phase of the receiving live clock.
\end{quote}
When this phase constraint is met for a channel between a transmitting live
clock and a receiving live clock, we say the receiving live clock is
\emph{logically synchronized} to the transmitting live clock.  Logical
synchronization is analogous to the coordination between neighboring people in
a bucket brigade, or that between players tossing a ball back and forth, where
the arrival of the ball must be within a player's `phase of catching'.  In this
way the notion of a channel is expanded to include the clock readings that
indicate phases of signal arrivals that have to be controlled in order for the
logical synchronization of the channel to be maintained.  (While in many cases
the integers in clock readings that count cycles can be definitely specified,
the phases are never exactly predictable.)  We model the phase of writing at
which a live clock can receive a character as corresponding to
\begin{equation}\label{eq:ls}
  |\phi| < (1-\eta)/2,
\end{equation}
where $\eta$ (with 0 $< \eta < 1$) is a phase interval that makes room for
reading.

Logically synchronizing a channel means bringing about the condition
(\ref{eq:ls}) on phases at which signals arrive.  Once logical synchronization
is acquired, maintaining it typically requires more or less continually
adjusting the rates of ticking and the acceleration of one or both of the live
clocks, in order to steer the phases of arriving characters toward a suitable
aiming point, say some $\phi_0$.  In the simplest case, this aiming point
$\phi_0$ is 0.  When a live clock receives signals over more than one channel,
it measures its own clock readings at all the receptions and takes all the
phase deviations into account to steer its own tick rate and acceleration.  The
clock readings made for the purpose of maintaining logical synchronization are
not interpreted as indicating ``the time''; instead, like the readings of a
clock that is being adjusted by a clockmaker, they indicate the amount by which
the live clock needs to be adjusted.

Relations among readings of live clocks that contribute to an aiming
point for a network include  what we call echo counts, closely related to
distances defined by radar:
  \begin{quote}
  \textbf{Definition of echo count:} Suppose that at its reading $m.0$
  a live clock $A$ transmits a signal at to a live clock $B$,
  and the first signal that $B$ can transmit back to $A$ after
  receiving $A$'s signal reaches $A$ at $m'.\phi'$; then the quantity
  $m'.\phi'- m.0$ will be called the echo count $\Delta_{ABA}$ at $m$.
  \end{quote}
The need for logical synchronization of the channels strongly constrains the
design of a network of live clocks.  In Sec.\ \ref{sec:4}  we will see how the
ticks of the clocks are allowable only within intersections of ``stripes in
spacetime'' \cite{aop14}.

Steering toward an aiming point involving echo counts depends on one live clock
receiving signals that convey readings of other live clocks.  The possibility
of wireless communication requires that such a signal from a live clock $A$ to
a live clock $B$ carry an identifier of $A$, so that the receiver $B$ can tell
the source of the signal.  With these identifiers, the channels of a network of
live clocks correspond to directed edges of a graph, leading to a data
structure for readings of live clocks of a network illustrated in
[\citenum{aop14}] and discussed at more length in terms of marked graphs and
Petri nets in [\citenum{1639}].  Such data structures for a network can reside
in the memory of a live clock, so the live clock has a picture of the network
in which it participates.

\section{Numerical communication as a frame of reference}\label{sec:4}
With unpredictability established in Sec.\ \ref{sec:2} and the consequent need
for feedback to support logical synchronization established in
Sec.\ \ref{sec:3}, we now show how networks of live clocks can offer 
reference patterns of motion against which to think about and measure the
motion of some object of interest.  The following subsections tell:
\begin{enumerate}
\item how the concept of a coordinate system as the prevalent reference for
  motion  depends on Einstein's synchronization criterion and thereby implicitly
  depends on a network of live clocks;
\item how, assuming a flat spacetime, the requirement for logical
  synchronization constrains a network of live clocks that approximates a
  Lorentz frame;
  \item how logical synchronization can be maintained in cases that preclude Einstein
  synchronization;
  \item how, in the theory of curved spacetime, there can be no
  large  rigid body and no all-purpose network of live clocks to realize
  a background to motion;
  \item how a network of live clocks serves as a tool to discover features of
    its unpredictable environment, including features of gravitation;
  \item how predictability meets unpredictabilty on a Turing tape.
\end{enumerate}

\subsection{Coordinate frames presuppose communication between `self-aware' clocks}
To see that a coordinate system presupposes a network of live clocks, recall
Einstein's definition of synchronization in special relativity.  Einstein
conceived of the location of an arbitrary event as a clock tick coincident with
the event.  This makes a spacetime of possible locations of events into a
potential for clock ticks, so that for any event one can think of a tick
coincident with it.  But the tick is significant only in relation to other
ticks.  By imagining a system of light signals propagating between imagined
\emph{proper} clocks, Einstein defined the synchronization of proper clocks
fixed to a non-rotating, rigid body in free fall (i.e., a Lorentz frame) and
co-defined ``time'' as the readings of such proper clocks, with the
implications that distance from proper clock $A$ to proper clock $B$ is
defined, as in radar, in terms of the duration at $A$ from the transmission of
a light signal to the return of its echo from $B$.  (If we imagine proper
clocks as live clocks, this difference in readings of clock $A$ is an echo
count.)  Specifically, according to Einstein's definition of the
synchronization of proper clocks fixed to a Lorentz frame \cite{einstein05},
clock $B$ is \emph{synchronous} to clock $A$ if at any $A$-reading $t_A$, $A$
could send a signal reaching $B$ at $B$-reading $t_B$, such that an echo from
$B$ would reach $A$ at $A$-reading $t'_A$, satisfying the criterion
\begin{equation}\label{eq:es}
t_B=\half(t_A+t'_A).  
\end{equation}
A Lorentz coordinate system presupposes Einstein's synchronization and thereby
presupposes the possibility of proper clocks that exchange light signals---an
abstraction of a grid-like network of live clocks and their signals, in which
every pair of live clocks exhibits constant echo counts and a constant Einstein
synchronization relation with its neighbors.

The clock readings along the history of a moving object constitute a
description of its motion, but how does that description become knowable to the
investigators?  The concept of a Lorentz coordinate system obscures the
information processing needed for clock readings to become known to
investigators.  In contrast, live clocks allow for the local recording of their
readings and the subsequent transmission of readings from one live clock to
another.  They also give expression to the necessity, stemming from quantum
theory, to respond to unpredictable deviations from desired relations.

Although special relativity imagines proper clocks free of drift, quantum
theory asserts an irreducible drift in tick rates and echo counts \cite{aop14},
so that the readings of nearby, unadjusted clocks wander apart without bound
unless their ticking is adjusted.  When we recognize the inescapability of this
drift, the Einstein synchronization relations (\ref{eq:es}) and the echo counts
can no longer be taken as externally supplied facts, but instead work as aiming
points chosen by investigators, toward which to steer the operation of a
network of live clocks.  This steering, if it can be accomplished, must depend
on feedback of unpredictable measured deviations from the aiming point into
the steering of the tick rates and the accelerations of the clocks of a
network.

To be steerable on the basis of measured deviations, the clocks must act as
live clocks.  Furthermore, feedback that steers toward Einstein's
synchronization requires the communication of numerals from one live clock to
another.  For example, consider a network of live clocks, subject to
unpredictable drift, in which each live clock aims to satisfy an aiming point
expressed as a set of relations among readings of itself and neighboring live
clocks.  Suppose the aiming point is comprised of the Eq.\ (\ref{eq:es}) along
with constant echo counts for selected channels to nearby live clocks.  Because
of drift, measured clock readings deviate from the relations that constitute
the aiming point.  In response to the deviations from the aiming point, each
live clock promptly adjusts its tick rate and acceleration.  A live clock
cannot wait for an external intelligence to compute the deviations exhibited by
measured, unpredictable clock readings; it must itself compute the relevant
deviations and its own response to these.  For example, a clock $B$ computes a
deviation that involves the difference between its own reading at its reception
of a signal from a nearby clock $A$ and the reading of $A$ at the transmission
of that signal.  Thus $B$ can know how it deviates from being
Einstein-synchronous with $A$ only if readings of $A$'s clock are communicated
to $B$.  It is not enough for a light signal from $A$ to reach $B$; in addition
the light signal has to convey the numerical reading of $A$ at the transmission
of that signal.  Hence steering toward Einstein's synchronization criterion of
Eq.\ (\ref{eq:es}) requires the communication of numbers, which requires
logical synchronization.

\subsection{Constraints of logical synchronization limit realizations of coordinates}
The requirement for logical synchronization constrains the arrangements of
clocks that can satisfy Einstein's synchronization criterion. To show this, it
is instructive to consider the case of ideal logical synchronization in which
all phases at signal receptions are zero.  Take the theoretical case of 8 live
clocks located at the corners of a cube in a flat or conformally flat
spacetime.  Let each edge of the cube correspond to logically synchronized
channels in both directions, all with zero phases.  In that case channels
across the face diagonals cannot have zero phases, simply because the ratio of
the length of a face diagonal to an edge is irrational.  But if the cube is
replaced by a rectangular brick, the brick can be chosen as an Euler brick for
which edges and face diagonals can both have integer lengths \cite{brick},
allowing a pattern of signaling having receptive phases for signals along edges
of the brick and along its face diagonals to all be zero.  In effect the
requirement for logical synchronization puts stripes on spacetime for allowable
configurations networks of live clocks \cite{aop14}.  Nonetheless, within any
theory that assumes a flat or conformally flat spacetime, a three dimensional
grid of live clocks exchanging signals along edges and face diagonals of Euler
bricks serves as a universally applicable pattern of motion for a coordinate
system.  We will soon see a difference in this regard between a conformally
flat spacetime and a generic curved spacetime.

\subsection{Logical synchronization where Einstein synchronization fails}
While realizing Einstein's synchronization requires logical synchronization
among pairs of live clocks, the converse does not hold: cases of logically
synchronized networks can be demonstrated for which Einstein synchronization is
impossible.  These cases include live clocks in relative motion, subject to the
Doppler effect, live clocks on a rotating platform, subject to the Sagnac
effect, and live clocks in the presence of gravitational fields. For this and
other reasons it becomes an interesting scientific topic to explore possible
patterns of signal exchange among logically synchronized live clocks.  The
topic includes the application of logical synchronization to the study of
gravitation.

\subsection{Live clocks under the assumption of spacetime curvature}

Bringing hypotheses of one or another curved metric tensor field into
consideration reveals a conceptual challenge to the definition and measurement
of motion.  As amply confirmed by experiments with space vehicles, by
astronomical observations, and by experience with the Global Positioning System
(GPS), the theory of general relativity asserts that the flat metric tensor
field of special relativity and its concomitant realization by a pattern of
light signals between clocks can only be approximated over a region the size of
which must decrease as the instability of clocks decreases.  If generic
curvature is significant, there can be no arbitrarily fine, three-dimensional
grid of signal-exchanging live clocks that pairwise satisfy Eq.\ (\ref{eq:es}),
so that the universally applicable pattern of motion for a coordinate system
noted for flat spacetime is ruled out \cite{perlick}.  This effect of generic
spacetime curvature is a consequence of general relativity alone without
invoking logical synchronization.  Because the concept of a rigid body hinges
on maintaining Einstein synchronization among the proper clocks thought of as
elements of that body, it is apparent that the theory of general relativity is incompatible
with any exactly rigid body.

\subsubsection{How to accept ``no rigid bodies''}
A pattern of motion taken as a tool for determining my own motion serves as a
tool for navigation, perhaps best appreciated in circumstances of its absence,
as when, perhaps on a hike, I am lost.  The thought of being lost is scary.  It
is soothing to imagine a flat coordinate system, relative to which the distant
and the immediate are brought into relation, the ground under my feet stays
put, and navigation would be straightforward.  But if the earth shifts, as at
some level of precision it always does, how is one to think?  And at levels of
precision made possible by modern clocks, there can be no rigid body on which
to stand, nor can there be a fixed background of communication.  It becomes
necessary to search for a background pattern and to adjust it if increased
demands for precision or unforeseen circumstances make the sought pattern
unrealizable.  If needs to adjust are in the cards, it is perhaps better to be
nimble.  Recognizing unpredictability can be a first step toward that nimbleness.

\subsection{Logical synchronization opens up possibilities for discovery}
While we have spoken of spacetime coordinates, the concept of a network of
live clocks as a reference for motion makes no assumption of a spacetime
manifold or of a coordinate system.  To design an aiming point of desired
relations among clock readings for a network of live clocks, however, one needs
to invoke some hypothesis about the propagation of signals from one live clock
to another live clock.  What makes a workable hypothesis depends on
circumstances, for example whether the signal is conveyed by light in vacuum,
by light in some medium such as optical fiber, or by nerve pulses, etc.

As a first specific case appropriate to theoretical physics, consider live
clocks on space vehicles linked in a communications network for an experiment
aimed at discovering features of gravitation that affect the network.  Suppose
the experiment employs a theory of of signals as light-like geodesics in some
specified spacetime manifold.  Then the theory of propagations of signals among
live clocks depends on the curvature of a metric tensor field hypothesized for
the spacetime manifold. To maintain logical synchronization, the live clocks
employ control equations that depend not only on the investigators' basic
assumption of a (possibly curved) spacetime manifold, but also on an additional
provisional hypothesis of a particular metric tensor field, from which they
arrive at a model of signal propagation, necessary to designing an aiming point
specified in terms of relations among readings of live clocks at the
transmission and the reception of light signals.  The control equations depend
on the provisional metric tensor field, and, together with that metric tensor
field, are subject to revision when difficulties in maintaining logical
synchronization are encountered.

For example, suppose the investigators guess that their space vehicles behave
as if they were in a region of spacetime that has a metric that is flat to
within the tolerances they can achieve with their live clocks.  To test this
guess, the investigators choose relations among clock readings as an aiming
point under the provisional hypothesis of a flat spacetime.  If, counter to
their guess, the live clocks deviate too much from the aiming point, that
deviation indicates the effect of a Weyl curvature of the spacetime.

In [\citenum{aop14}] a specific case of a cluster of 5 space vehicles, each
equipped with a live clock, is analyzed for its capacity to measure
unpredictable changes in gravitation by utilizing feedback to adjust clock
rates and clock accelerations to generate a background motion maintained by
computer-mediated feedback in response to unpredictable events.  Gravitational
effects are then detectable as they influence the channels found to be
implementable.  The investigation of gravitational effects thus involves a
cycle of hypothesizing a metric tensor field and testing the implications of
that hypothesis for clock readings against those recorded by the 5 live clocks.

 Under hypotheses defining the second as a measuring unit in the
International System of Units (SI) \cite{aop14}, and with sufficiently
 stable live clocks, such a spaceborne network can potentially measure ripples
 in gravitation, something already pursued by other but related means by the
 Laser Interferometer Gravitational Wave Observatory (LIGO) \cite{ligo}.  We
 suspect that LIGO in fact illustrates how the use of feedback that responds to
 unpredictable events allows a precision in the background of measurement
 otherwise unattainable, and hence opens experimental inquiry into effects
 otherwise invisible.  The complexity of LIGO, however, led us to point to the
 above arrangement of devices that, although likely difficult to implement, is
 conceptually simpler in its use of feedback that responds to the
 unpredictable.

\subsection{Predictability meets unpredictabilty on a Turing tape }
As represented in the memory of a live clock a variable can be thought of as a
stretch of a Turing tape. A live clock engaged in a feedback loop houses separate
variables---separate stretches of its Turing tape---for an aiming
point derived from a hypothesis as distinct from a measured, unpredictable
deviation from that aiming point.  While both variables for aiming points and
variables for deviations can experience changes in their values, typically the
rate of change is relatively high for a deviation and low for an aiming point.

Important to logical synchronizaton is the distinction between a count of
cycles recorded on tape at the receipt of a signal over a logically
synchronized channel and the recorded phase of the cycle at which the signal
arrives.  The count is so to speak a ``public number'' that ought to be the
same if the live clock were interchanged with another live clock; the phase
however, is essentially an analog rather than a digital business; the phase is
idiosyncratic in the sense that there must be a tolerance within which it would
differ in one live clock were substituted for another.  Idiosyncratic phases
are indispensable to the control that permits the logically coherent
transmission of numerals from one live clock to another, so that a "7" on a
tape of live clock $A$ becomes a "7" on live clock $B$, and arithmetic
functions across a communications channel.

When we set aside the concept of a coordinate system in favor
of a network of live clocks, with its feedback loops that act on its patterns of
signal exchange in response to unpredictable influences, do we have anything
to hold onto in place of the fixed relations that are most vivid in the
rigid body that anchors coordinates in special relativity?  What a network
of live clocks offers is both a communications facility and a conceptual
framework to think about connections among entities of interest.  This
conceptual framework enriches the notion of a coordinate system in
circumstances in which a coordinate system is appropriate.  It also serves in
situations in which unpredictability makes a coordinate system unavailable.
What one holds onto are the records of the communications channels and of the
unpredictable adjustments necessary to their maintenance found in the memories
of live clocks.

\section{Discussion}\label{sec:5}
This paper builds on the more detailed analysis of unpredictability in clocks
reported in ref.\ [\citenum{aop14}], in which live clocks are called ``open
machines.''

Guesswork as a necessity has a long history of proponents.  For example,
Einstein endorses Hume's showing that experimental evidence can never establish
a causal connection, and Einstein goes further to say that all our concepts
depend on ``freie Sch\"opfungen des Denkens''---free creations of
thought---i.e. guesses \cite{weltbild}.  The virtue of the proofs noted in
Sec.\ \ref{sec:2} in not to bring a new thought, but to demonstrate it from
within physics.

For computation to work over a communications channel, it is necessary for the
steps of computation to be adjusted in response to unpredictable, idiosyncratic
phases of signal arrivals over logically synchronized channels.  Channels in
networks get put into operation, are maintained for a while, and then perish.
The memory of a live clock can house a picture of the network at some moment,
and this picture may or may not agree with a picture held in the memory of
another live clock; indeed, communications delays can make two such pictures
held in separated memories incommensurate.  The coming into operation and the
perishing of channels are almost entirely outside the scope of this report: the
maintenance of logically synchronized channels alone, while only the simplest
part of the story, shows how, by invoking guesses of how to respond to
unpredictable measured detections, live clocks step their computations in a way
that allows the spread of arithmetic across a network.  Attention to the
networks of live clocks as references for motion shows ways in which the
unpredictable and the calculated work in combination during scientific
discovery, and how that ``working together'' rules out the view that what is
discovered is out there independent of we who, with our dependence on
guesswork, go about looking for it.

Contact with the physical world is reflected in records held in the memories of
computers, seen here as live clocks.  Such memories include traditional
libraries holding the extant physics literature.  The records in memories
distributed over networks of computers depend on assumptions, as does the use
of these records in making predictions and in designing experiments.  The
assumptions that reside in memories and are applied, for example in managing
logical synchronization, vary in what one might call depth.  At root are
community-wide assumptions, including conventions of language and logic.  From
these assumptions, that investigators are slow to change, there range more
flexible assumptions, up to those that an individual can make today and drop
tomorrow.  Under the impact of responses to surprises to which assumptions are
vulnerable, this tangle of axioms, postulates, hypothes and whatever else we
call it that comes into our heads without a logical justification, which we
call a ``tree of assumptions,'' evolves unpredictably; it can be glimpsed by
any person only in a small part, and a person's capacity to notice an
assumption depends on what other assumptions that person carries or rejects at
the moment.

Maintaining logical synchronization among live clocks that contribute to a
reference background of motion must draw on a tree of assumptions that puts a
gap between evidence and its explanations.  The numbers that come as evidence
into a network affect the network and are distinct from explanations in terms
of particles and fields, also numeric structures, which presuppose a coordinate
system.  Pretending that particles or fields can enter records directly as
evidence obscures the gap between evidence and explanation.  Accepting the
unpredictability in a reference background that stems from the guesswork
intrinsic to the tree of assumptions engenders a deeper sense of order,
resident in memories of live clocks, and operative in situations beyond those
for which the notion of a coordinate system finds application.

Accepting unpredictability in the backgrounds against which all else is seen
takes a certain courage.  The reward is the opportunity to learn to work with
unpredictability in physics, thereby entering a world in which the joy of an
answer can be the new question it enables.

\section*{Acknowledgment}
We thank Prof. Tai Tsun Wu for discussions of the topic of discovery.

\end{document}